\documentclass[prb,twocolumn,amsmath,amssymb,letterpaper,floatfix,superscriptaddress]{revtex4}
\usepackage{graphicx}
\usepackage{color}
\usepackage{amsfonts}
\usepackage{bm}
\def\tHp{\tau_{\rm Hp}}

\def\SHp{S_{\rm Hp}}
\def\RHp{Re_{\rm Hp}}

\def\vpo{v_{\vartheta 0}}
\def\qMin{q_{\rm min}}
\def\qRes{q_{\rm s}}

\def\rsi{r_{{\rm s}i}}

\def\rsa{r_{\rm s1}}
\def\rsb{r_{\rm s2}}

\def\rMin{r_{\rm min}}

\def\mMax{m_{\rm max}}

\def\mPeak{m_{\rm peak}}

\def\fLin{\omega_{\rm lin}}

\def\flPsi{\widetilde{\psi}}
\def\flPhi{\widetilde{\phi}}

\def\gLin{\gamma_{\rm lin}}

\def\O{\mathcal{O}}

\definecolor{gray}{rgb}{0.5,0.5,0.5}
\definecolor{dred}{rgb}{0.5,0.0,0.0}
\definecolor{dgreen}{rgb}{0.0,0.5,0.0}
\definecolor{dblue}{rgb}{0.0,0.0,0.5}

\begin{document}

\title{Large-mode-number magnetohydrodynamic instability driven by sheared flows \\ in a tokamak plasma with reversed central shear}

\author{Andreas~Bierwage}
\email{abierwag@uci.edu}
\altaffiliation[Present address: ]{Department of Physics and Astronomy, University of California, Irvine, CA 92697, U.S.A.}
\affiliation{Max-Planck-Institut f\"{u}r Plasmaphysik, EURATOM Association, D-85748 Garching, Germany}
\author{Qingquan~Yu}
\affiliation{Max-Planck-Institut f\"{u}r Plasmaphysik, EURATOM Association, D-85748 Garching, Germany}
\author{Sibylle~G\"{u}nter}
\affiliation{Max-Planck-Institut f\"{u}r Plasmaphysik, EURATOM Association, D-85748 Garching, Germany}

\date{\today}

\begin{abstract}
The effect of a narrow sub-Alfv\'{e}nic shear flow layer near the minimum $\qMin$ of the tokamak safety factor profile in a configuration with reversed central shear is analyzed. Sufficiently strong velocity shear gives rise to a broad spectrum of fast growing Kelvin-Helmholtz (KH)-like ideal magnetohydrodynamic (MHD) modes with dominant mode numbers $m,n\sim 10$. Nonlinear simulations with finite resistivity show magnetic reconnection near ripples caused by KH-like vortices, the formation of turbulent structures, and a flattening of the flow profile. The KH modes are compared to double tearing modes (DTM) which dominate at lower shearing rates. The possible application of these results in tokamaks with internal transport barrier is discussed.
\end{abstract}


\maketitle

\thispagestyle{empty}

\begin{figure}
[tb]
\centering
\includegraphics[height=5.6cm]
{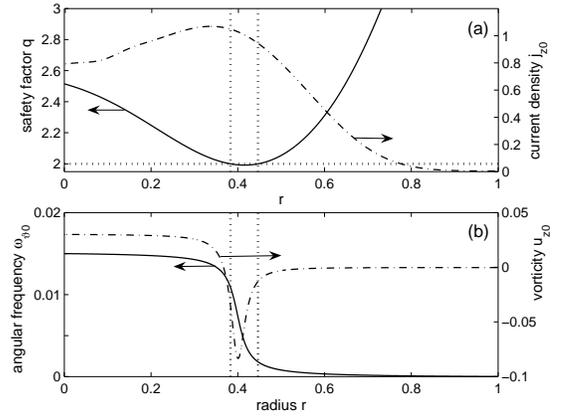}
\caption{
Equilibrium profiles of (a) the safety factor $q$ and current density $j_{z0}$, and (b) the angular frequency $\omega_{\vartheta 0}$ and vorticity $u_{z0}$. $\qRes = 2$ resonances are indicated by dotted lines.}
\label{fig:equlib}%
\end{figure}

Shear flows can give rise to the well-known Kelvin-Helmholtz instability (KHI) which, in magnetohydrodynamic (MHD) systems, has to compete with field line tension, magnetic pressure and magnetic shear.\cite{Chandrasekhar} Regions in the plasma which are particularly susceptible to KHI are therefore those where it can couple with MHD instabilities. A system where such coupling is expected to play an important role, and which motivated the present study, is the reversed-shear (RS) advanced tokamak configuration, which is a strong candidate for achieving self-sustained nuclear fusion conditions. It has a non-monotonic toroidal current profile [Fig.\ref{fig:equlib}(a)] and the magnetic shear, $s = (r/q){\rm d}q/{\rm d}r$, changes its sign at some radius $\rMin$. Here $r$ is the radial coordinate (minor radius of the torus), and $q(r)$ is the safety factor profile measuring the field line pitch [Fig.\ref{fig:equlib}(a)]. In the region with negative shear ($r < \rMin$) pressure-driven ideal MHD ballooning modes are stable, so that higher pressures can be achieved in the core. Strongly sheared poloidal flows may further improve confinement in RS tokamaks. There is evidence that such zonal flows play an important role in the formation of an internal transport barrier (ITB) near $\rMin$.\cite{Bell98, Connor04a} MHD instabilities are often observed in ITB discharges.\cite{Levinton98, Guenter00} While there is evidence that certain ITB equilibria with strong pressure gradient in the region of weakly positive shear may be stable to ideal MHD ballooning modes,\cite{Connor04b} low shear around $\qMin$ may allow pressure-driven infernal modes to develop.\cite{Manickam87, Ganesh05} Recently, non-curvature-driven instabilities have also attracted attention.\cite{Rogers05, Tatsuno06} The destabilization of an ideal electrostatic KHI in a region with low magnetic shear was demonstrated in Ref.~\onlinecite{Idomura00}. KHI coupled to parallel magnetic perturbations for $s=0$ was considered in Ref.~\onlinecite{Miura01}.

The scenario considered in this Letter consists of a radially localized shear flow layer near $\qMin$ in an equilibrium which is also unstable to a broad spectrum of current-driven resistive double tearing modes (DTM).\cite{Bierwage05a, Bierwage05b} This may occur whenever $\qMin$ passes through a low-rational value.\cite{Levinton98, Guenter00} Shear flow is known to stabilize DTMs through its tendency to decouple the resonances.\cite{Ofman92, Dewar93} In the nonlinear regime, locking between finite-size magnetic islands and flattening of the velocity gradient may destabilize DTMs.\cite{Persson94, Yu97} Thus, a dynamic interaction between DTMs and shear flows is expected.\cite{Guenter00}

In this Letter, we describe in cylindrical geometry how a broad spectrum of linear DTMs is stabilized by increasing velocity shear and show that, above a certain threshold, fast growing electromagnetic KH-like modes with dominant high mode numbers $m \sim 10$ appear. The linear instability of both DTMs and KH-like modes is analyzed and first nonlinear simulation results are presented for a case with finite resistivity. The latter shows magnetic reconnection, the formation of turbulent structures, and a flattening of the flow profile.

\begin{figure*}
[tb]
\centering
\includegraphics[height=5.8cm]
{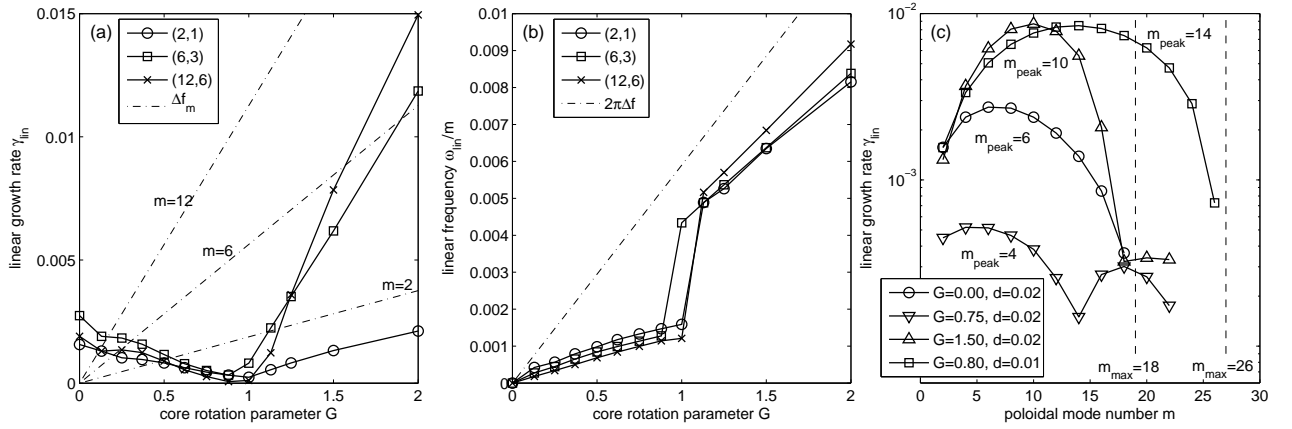}
\caption{(a) Linear growth rate $\gLin$ and (b) rotation frequency $\fLin/ m$ for the modes $(m,n)=(2,1)$, $(6,3)$ and $(12,6)$, in dependence of the core rotation parameter $G$. The effective and specific shearing rates, $\Delta f_m$ and $\Delta f$, are plotted as dash-dotted lines. In (c) spectra of linear growth rates $\gLin(m)$ are shown for three cases with $d=0.02$: $G=0$ stationary, $G=0.75$ stabilized, $G=1.5$ destabilized. A spectrum obtained with a narrower shear layer is also shown ($d = 0.01$, $G = 0.8$).}
\label{fig:scans}%
\end{figure*}

For simplicity, we use the well-known reduced magnetohydrodynamic (RMHD) model in cylindrical geometry $(r,\vartheta,z)$.\cite{Strauss76} In the limit of zero pressure, the RMHD equations for the magnetic flux $\psi$ and the vorticity $u_z$ are
\begin{eqnarray}
\partial_t\psi + {\bf v}\cdot\bm{\nabla}\psi &=& -\eta j_z
\label{eq:rmhd1}
\\
\rho_{\rm m}\left(\partial_t u_z + {\bf v}\cdot\bm{\nabla} u_z\right) &=& {\bf B}\cdot\bm{\nabla} j_z + \nu \nabla_\perp^2 u_z.
\label{eq:rmhd2}
\end{eqnarray}

\noindent Here, $\rho_{\rm m}$ is the mass density, $\eta$ the resistivity and $\nu$ the viscosity. Current density $j_z$ and vorticity $u_z$ are related to $\psi$ and $\phi$ through $\mu_0 j_z = -\nabla_\perp^2\psi$ and $u_z = \nabla_\perp^2\phi$, respectively. In the following, time $t$ is measured in units of the poloidal Alfv\'{e}n time $\tHp = \sqrt{\mu_0 \rho_{\rm m}} a/B_0$ (with $B_0$ being the strong axial magnetic field) and the radial coordinate $r$ is normalized by the minor radius of the plasma, $a$. Resistive dissipation is measured by the magnetic Reynolds number $\SHp = \tau_\eta/\tHp$ (with $\tau_\eta = a^2\mu_0/\eta$ being the resistive diffusion time) and viscous damping by the kinematic Reynolds number $\RHp = a^2/\nu\tHp$. The perturbed fields are Fourier transformed as $\widetilde{h} = \sum_{m,n} \widetilde{h}_{m,n}(r)\exp(im\vartheta - inz/R_0)$ (with $R_0$ being the major radius of the tokamak), where $m/n = q_{\rm s} = 2$ (single-helicity perturbation). The model and numerical methods are described in detail in Refs.~\onlinecite{Bierwage05b} and \onlinecite{Bierwage06a}.

The equilibrium configuration used is shown in Fig.~\ref{fig:equlib}. The safety factor has two $\qRes = 2$ resonant surfaces located a small distance $D_{12} = \rsb-\rsa = 0.066$ apart ($\rsa = 0.381$, $\rsb = 0.447$). The magnetic shear at these surfaces is $s_1 = -0.10$ and $s_2 = 0.12$. The poloidal velocity profile $\vpo$ has a shear layer near $\rMin$ and is given by
\begin{equation}
\omega_{\vartheta 0} = \vpo/r = \Omega_0 G \left[C_1 - \tan^{-1}\left((r-r_0)/d\right)\right]/C_0,
\label{eq:vpo}
\end{equation}

\noindent with $C_1 = \tan^{-1}[(1-r_0)/d]$ and $C_0 = C_1 - \tan^{-1}[(-r_0)/d]$. The standard case for $\omega_{\vartheta 0}$ is shown in Fig.~\ref{fig:equlib}(b). The shear layer is located near the inner resonance ($r_0 = 0.4$) and has a width ($2d = 0.04$) comparable to $D_{12}$. The parameter $G$ controls core rotation and thus the shearing rate between the resonances. We define the specific shearing rate as $\Delta f = (\omega_{02}-\omega_{01})/2\pi$ [with $\omega_{0i} = \omega_{\vartheta 0}(\rsi)$]. The effective shearing rate for the mode $(m,n)$ is $\Delta f_m = m\Delta f$. Since the equilibrium flow is not included self-consistently, only sub-Alfv\'{e}nic velocities are considered ($\Omega_0 = 0.01$, $|G| \leq 2$). The characteristics of the shear flow are similar to those reported in Ref.~\onlinecite{Bell98}. Both the static MHD equilibrium and the shear flow are separately unstable: the former to resistive DTMs\cite{Bierwage05b} and the latter to the ideal KHI:\cite{Chandrasekhar} the Rayleigh criterion for KHI is fulfilled [$u'_{z0}(r_0) = 0$] and the existence of an instability threshold $\mMax$, is demonstrated shortly.

For the shear-flow-driven modes described in this Letter the vorticity gradient term $\rho_{\rm m}(im/r)\widetilde{v}_r u_{z0}'$ ($u_{z0}' = {\rm d}u_{z0}/{\rm d r}$) in Eq.~(\ref{eq:rmhd2}) is crucial. This term is known to be the origin of KHI.\cite{Chandrasekhar} In accordance with the criterion given by Eq.~(23) in Ref.~\onlinecite{Tatsuno06}, the presence of resonant surfaces is found to be crucial for instability. We find that low-$m$ modes even require \emph{two} nearby resonances and that multiple branches of unstable modes (ideal and resistive) can be destabilized simultaneously.

\begin{figure}
[tb]
\centering
\includegraphics[height=6.2cm]
{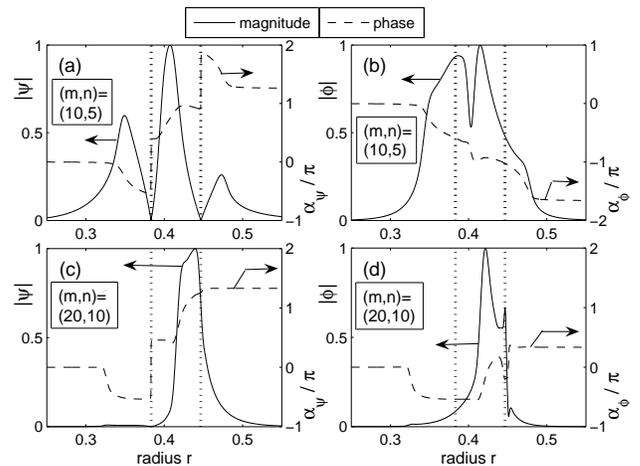}
\caption{Linear eigenmode structures for $(m,n)=(10,5)$ [(a) and (b)] and $(20,10)$ [(c) and (d)]. The flow parameters are $r_0=0.4$, $d=0.02$ and $G=1.50$. Flux $\flPsi$ and potential $\flPhi$ are decomposed into magnitude and complex phases as $\psi=|\psi|e^{i\alpha_\psi}$. Vertical dotted lines indicate resonant surfaces.}
\label{fig:mstruc}%
\end{figure}

Results obtained by solving the linearized RMHD equations as an initial-value problem are presented in Figs.~\ref{fig:scans} and \ref{fig:mstruc}. In Fig.~\ref{fig:scans}(a) the linear growth rates $\gLin$ and in Fig.~\ref{fig:scans}(b) the linear frequencies $\fLin$ of several modes are plotted as functions of $G$. Spectra of unstable modes are shown in Fig.~\ref{fig:scans}(c) for several values of $G$. The linear mode structures are shown in Fig.~\ref{fig:mstruc}. The data were obtained with $\SHp = 10^7$ and $\RHp = 10^{11}$, so the viscosity effect is negligible.\cite{Bierwage05b}

The $\gLin(G)$ curves in Fig.~\ref{fig:scans}(a) are symmetric around $G=0$ (no flow), so only $G\geq 0$ is considered. Since the distance $D_{12}$ is small, the resonances couple strongly and a broad spectrum of resistive DTMs with dominant high poloidal and toroidal mode numbers $m$ and $n$ are found for $G=0$ (cf. Refs.~\onlinecite{Bierwage05a, Bierwage05b}). In the range $0 < G \lesssim 1.0$, the effect of the shear flow on DTMs is primarily a stabilizing one. However, it can be seen in Fig.~\ref{fig:scans}(a) that this stabilization may stagnate and even reverse. For instance, the growth rate of $(m,n) = (12,6)$ mode rises briefly around $G\sim 0.2$ before descending further. The origin of this behavior is not yet understood. The growth rates reach a minimum around $G_{\rm min} = 1.0$ and increase for $G > 1.0$ almost linearly with $G$. When a smaller velocity gradient ($\propto G/d$) is used, or when the inter-resonance distance $D_{12}$ is increased, a larger value is obtained for $G_{\rm min}$ (and \textit{vice versa}). Without resonant surfaces ($\qMin > 2$) no unstable modes with helicity $m/n=2$ are found.

The frequencies $\fLin(G)$ in Fig.~\ref{fig:scans}(b) undergo a shift around $G \approx G_{\rm min}$. For $G < G_{\rm min}$ the rotation at the outer resonance determines the frequency of the mode. This is typical for DTMs (since typically $s_1 < s_2$) and indicates that the current drive dominates in this regime. For $G > G_{\rm min}$ the frequency equals that at the inner resonance, which is closer to the shear layer. Hence, $G = G_{\rm min}$ is identified as the transition point between predominantly current-driven and shear-flow-driven modes. With a shear layer closer to the outer resonance there is no frequency shift.

A comparison between the growth rates and the effective shearing rates in Fig.~\ref{fig:scans}(a) shows that $\gLin(G)$ increases with $G$ at least as fast as $\Delta f_m$. This suggests that some of the coupling between the resonances, which is important for DTMs, may pertain even for $G > G_{\rm min}$. For the $(2,1)$ mode we have observed this effect by comparing the results of the RS configuration with results obtained with a monotonic $q$ profile with a single $\qRes = 2$ resonant surface and otherwise similar conditions: in the former case (RS) the shear-flow-driven $(2,1)$ mode is unstable [cf.~Fig.~\ref{fig:scans}(a)], in the latter case (monotonic $q$) it is stable. In contrast, higher-$m$ modes such as $(6,3)$ and $(12,6)$ have similar growth rates in both cases, so for these the coupling between the resonances does not seem to play a crucial role.

The full spectra of linear growth rates $\gLin(m)$ are shown in Fig.~\ref{fig:scans}(c). For the same shear layer width as in Figs.~\ref{fig:scans}(a) and (b) (half width $d = 0.02$), the spectra are plotted for three values of $G$: no shear flow ($G=0$), stabilized ($G=0.75$), and destabilized ($G=1.5$). The dominance of modes with $m>2$ modes can be clearly seen. The DTM spectrum ($G=0$) has $\mPeak = 6$ as its fastest growing mode and the last unstable mode is $\mMax = 18$. In the stabilized case ($G=0.75$) $\mPeak$ has decreased to $4$. In addition, a second branch of weakly unstable modes is present, with $\mPeak = 18$. For $G=0.75$ the frequencies of modes on both branches are determined by the rotation of the outer resonance. In the destabilized case ($G=1.5$), the modes on the dominant branch around $\mPeak = 10$ grow rapidly and their frequencies are up-shifted, implying rotation with the inner resonance. The growth rates on the secondary branch are almost the same as for $G=0.75$ and the frequencies are still determined by the rotation of the outer resonance.

In Fig.~\ref{fig:scans}(c) a case with a smaller shear layer width $d=0.01$ is shown as well. Compared to the $d=0.02$ case, here the destabilization of shear-flow-driven modes has a lower threshold $G_{\rm min}$ and the spectrum is broader: $\mPeak = 14$ and $\mMax = 26$. Indeed, it is found that $\mMax$ is controlled by the parameter $d$, while $G$ seems to be irrelevant. This is similar to KHI which possesses an upper threshold for the wave number $k_\vartheta = m/r$ of unstable modes:\cite{Chandrasekhar} for $\gamma > 0$ it is required that $k_\vartheta d \lesssim \O(1)$. Here, for $d=0.02$ and $r_0=0.4$ the dominant branch has $\mMax = 18$ [Fig.~\ref{fig:scans}(c)], so that $k_\vartheta d = 0.9 \lesssim 1$. Modes on the secondary branch do not seem to be subject to the same threshold.

Calculations with magnetic Reynolds numbers in the range $10^7 \leq \SHp \leq 10^{15}$ reveal that shear-flow-driven modes ($G > G_{\rm min}$) on the dominant branch grow independently of $\SHp$. The ideal character of the instability is reflected in the linear eigenmode structures shown in Fig.~\ref{fig:mstruc}(a) and (b) for $(m,n)=(10,5)$: the flux perturbation switches signs at the resonances, as can be inferred from $|\flPsi|(\rsi) = 0$ and the phase shifts $\Delta\alpha(\rsi) = \pm\pi$ [Fig.~\ref{fig:mstruc}(a)]. The potential $\flPhi$ [Fig.~\ref{fig:mstruc}(b)] is non-zero at the resonances and its complex phase exhibits a jump at the inflection point of the velocity profile, where $u'_{z0} = 0$. Resistivity independent modes are also observed with a single resonant surface and otherwise similar conditions.

The typical structure of modes on the secondary branch is shown in Fig.~\ref{fig:mstruc}(c) and (d) for $(m,n)=(20,10)$. The perturbation is localized between the shear layer ($r_0$) and the outer resonance ($\rsb$). At $r=\rsb$, the flux $\flPsi$ is non-zero and the potential $\flPhi$ has a discontinuity, suggesting a tearing-type instability. This mode grows most rapidly for $\SHp \sim 10^7$, while the growth rate is reduced for larger and lower resistivity.

Further calculations show that the growth rates on the dominant branch are rather insensitive to the exact location of the shear layer between the resonances. This indicates that shear-flow-driven modes have a non-local character, in the sense that it is not sufficient to consider values of $\omega_{\vartheta 0}$ at the resonances only (cf. Ref.~\onlinecite{Tatsuno06}).

\begin{figure}
[tb]
\centering
\includegraphics[width=8.2cm]
{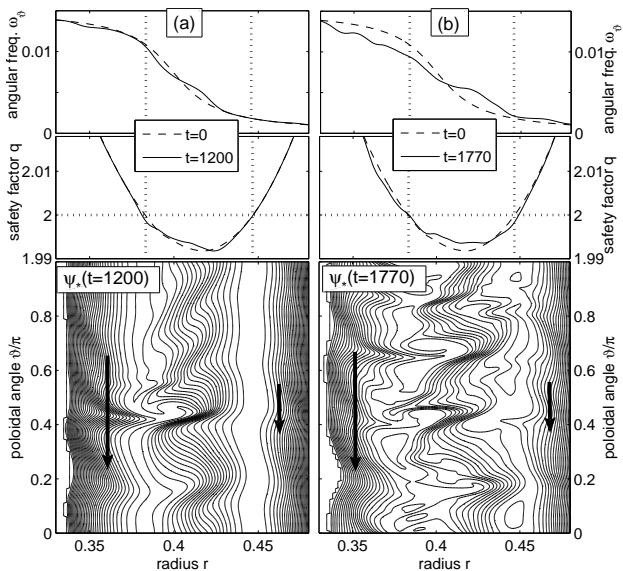}
\caption{Snapshots of nonlinear dynamics at (a) $t=1200$ and (b) $t=1770$. \emph{Top}: $\omega_\vartheta$ profiles. Dotted lines indicate the original resonant radii. \emph{Middle:} $q$ profiles. \emph{Bottom:} helical magnetic flux contours [$\psi_* = \psi + r^2/(2\qRes)$] in the $\vartheta$-$r$ plane.}
\label{fig:nonlin}%
\end{figure}

Finally, nonlinear simulation results are presented in Fig.~\ref{fig:nonlin} (parameters settings: $G=1.5$, $d=0.02$, $\SHp = 10^7$, $\RHp = 5\times 10^8$, single helicity $\qRes = 2$, 128 modes, dealiased). At the time of the first snapshot, Fig.~\ref{fig:nonlin}(a), the nonlinear mode coupling has reduced the velocity gradient around $r_0=0.4$, while the profile in the surroundings is steepened. The helical flux contours show islands around the inner resonance, with a dominant $m=2$ topology and high-$m$ ripples. In the X-point regions, poloidally localized KHI-like vortices (only one shown due to $\pi$-periodicity) can be observed. Subsequently, the $m=2$ islands undergo fragmentation and the structures become increasingly turbulent, as can be seen in the second snapshot in Fig.~\ref{fig:nonlin}(b).

Note that the shear-flow-induced perturbations extend over the entire inter-resonance region (and beyond). Thus, one may expect a coupling similar to that in DTMs. Indeed, Fig.~\ref{fig:nonlin}(b) shows island formation on both resonances while the central flux surfaces are still intact. Furthermore, in Fig.~\ref{fig:nonlin}(b) a flattening of the flow profile in a wide region can be observed, while $\qMin$ is still well below $\qRes = 2$. This shows that when resistivity is low, the primary saturation mechanism in the early nonlinear stage is likely to be the flattening of the flow profile.

The observed magnetic reconnection is due to finite plasma resistivity rather than turbulence effects. However, there are two instability driving mechanisms, the current gradient and the velocity gradient, and it is not clear how much each contributes at a given time. The observed flattening of the flow profile suggests that the current-gradient-driven instability (i.e., DTM) may play a more significant role in later stages. The study of the long-time evolution of this system requires turbulence simulations which go beyond the scope of the physical model and numerical code used here.

In summary, we have analyzed the effect of sheared flows in a RS tokamak configuration when $\qMin$ is below but close to a low-order rational value $\qRes = m/n$ and a pair of nearby resonant surfaces has formed. The growth rates of DTMs are reduced, although the spectrum remains peaked at $m > 2$ even with strong shear flow. For sufficiently strong velocity shear DTMs are replaced by a broad spectrum of shear-flow driven KH-like modes with dominant high $m \sim 10$. In the nonlinear regime and with finite resistivity, magnetic reconnection occurs near ripples caused by KH-like vortices. The tendency to develop turbulent structures in the inter-resonance region and flattening of the velocity profile is observed. Further calculations with experimental data as input are required in order to identify the role of KH-like modes in ITB discharges. Note also that since high-$m$ modes are involved the single-fluid MHD picture may be inaccurate. Our results motivate further investigations with more realistic models to understand the properties and effects of KH-like modes in tokamak plasma with ITB. Since high-$m$ modes are involved, we expect that similar instabilities may be found for other values of $\qRes$ such as $3/2$ or $7/3$.\cite{Bierwage05b}

A.~B. thanks the Max-Planck-Institut f\"{u}r Plasmaphysik in Garching for its support and hospitality, and gratefully acknowledges valuable discussions with S.~Hamaguchi, Z.~Lin, E.~Strumberger and T. Tatsuno.


\end{document}